\documentclass[prd,aps,nofootinbib,preprintnumbers,showpacs,showkeys]
{revtex4}
\usepackage{graphicx,epsf,amsfonts,amssymb,amsbsy}

\newcommand{\diag}{\mathop{\rm diag}\nolimits}
\newcommand{\e}{\mathop{\rm e}\nolimits}
\newcommand{\ts}{\textstyle}
\newcommand{\ds}{\displaystyle}
\newcommand{\ints}{\int\limits}

\newcommand{\eps}{\varepsilon}
\newcommand{\vev}[1]{\langle#1\rangle}
\newcommand{\Det}{\mathop{\rm Det}\nolimits}
\preprint{Preprint HU-EP-06/15}
\begin{document}
\large
\title{ \bf Quark and pion condensation in a chromomagnetic
background field}
\author{
D.~Ebert$^{1}$, K.~G.~Klimenko$^{2}$,  V.~Ch.~Zhukovsky$^{3}$, and 
A.~M.~Fedotov$^{3}$}
\affiliation{$^{1}$ Institut f\"ur Physik,
Humboldt-Universit\"at zu Berlin, 12489 Berlin, Germany}
\affiliation{$^{2}$ Institute
of High Energy Physics, 142281, Protvino, Moscow Region, Russia}
\affiliation{$^{3}$ Faculty of
Physics, Department of Theoretical Physics, Moscow State University,
119899, Moscow, Russia}
\newcommand{\be}{\begin{equation}}
\newcommand{\ee}{\end{equation}}
\begin{abstract}
The general features of quark and pion condensation in dense quark matter with
flavor asymmetry have been considered at finite temperature in the presence of a
chromomagnetic background field modelling the gluon condensate. In
particular, pion condensation in the case of a constant abelian
chromomagnetic field and zero temperature has been studied both
analytically and numerically. Under the influence 
of the chromomagnetic background field the effective potential of the
system is found to have a global minimum for a finite pion condensate even for small values
of the effective quark coupling constant. In the strong field limit,
an effective dimensional reduction has been found to take place. 

\end{abstract}


\maketitle
\renewcommand{\thefootnote}{\arabic{footnote}}
\setcounter{footnote}{0}
\setcounter{page}{1}
\section{ Introduction}

It is well-known that effective field theories with four-fermion
interaction
(the so-called Nambu -- Jona-Lasinio (NJL) models), which incorporate
the phenomenon of dynamical chiral symmetry breaking, are quite useful in
describing  low-energy hadronic processes (see
e.g. \cite{7,kunihiro} and references therein).
Since the NJL model displays the same symmetries as QCD,
it can be successfully used for simulating some of the QCD ground
state properties under the influence of external conditions such as
temperature, baryonic chemical potential, or even curved
space-time etc. \cite{kunihiro,buballa,odin,klev,klim} In particular,
the role of the NJL approach significantly increases, when
numerical lattice calculations are not admissible in QCD, like at nonzero
baryon density and in the 
presence of external gauge fields \cite{kl,ekvv}. 
In this way, it was observed in the framework of a (2+1)-dimensional
NJL model that an
arbitrary small external magnetic field induces the spontaneous
chiral symmetry breaking (CSB)
even at arbitrary weak interaction between fermions \cite{2}. Later,
it was shown that this phenomenon
(the so called magnetic catalysis effect) displays a universal
character
and can be explained basing on the dimensional
reduction mechanism \cite{gus,incera} (for a modern status of the 
magnetic catalysis effect and its applications in
different branches of physics, see the reviews \cite{gus2}).

As an effective theory for low energy QCD, the NJL model does
not contain any dynamical gluon fields. However, such a nonperturbative
feature of the real QCD vacuum, as the nonzero gluon condensate
$\vev{F^a_{\mu\nu}F^{a\mu\nu}}$ can be mimicked by external chromomagnetic fields.
In particular, for a 
QCD-motivated NJL model with gluon condensate (i.e. in the 
presence of an external chromomagnetic field) and finite
temperature, it was shown that a weak gluon condensate plays a
stabilizing role for the behavior of the constituent quark mass,
the quark condensate, meson masses and coupling constants for
varying temperature \cite{8}. Then, in a series of papers,
devoted to the NJL model with the gluon condensate, it was 
shown that an external chromomagnetic field, similar to the ordinary
magnetic field, 
serves as a catalyzing factor in the fermion mass generation
and dynamical breaking of chiral symmetry as well \cite{klim2}.
The basis for this phenomenon is also the effective reduction of the 
space dimensionality in the presence of strong external
chromomagnetic
fields, and this does not depend on the particular form of the field
configurations \cite {zheb}. 

At present time, it is well-established (see, e.g., 
\cite{saito}) that the gluon condensate is a very slowly decreasing
function of the baryon density (baryon chemical potential). So in
cold
quark matter, it is a nonzero quantity even at sufficiently large
baryon densities, which are expected to exist inside the neutron star
cores. Evidently, the 
consideration of the gluon condensate may change significantly the
generally
accepted picture of physical processes in the baryon matter. For
instance, according to the modern point of view, the 
color superconducting (CSC) phenomenon can be realized in neutron
star cores (see the reviews \cite{buballa}). However, according to 
\cite{agas}, the critical parameters of this phase transition
strongly depend on the value of the gluon condensate  (in this
case, only its chromomagnetic component survives).
Moreover, it turns out that even for a rather weak quark coupling,
different
external chromomagnetic field configurations induce the CSC
phenomenon \cite{ebklim}. Quite recently it was found that some
combinations of external chromomagnetic and ordinary magnetic
fields can penetrate into a bulk of the CSC medium and modify its
ground
state, producing a new type of color superconductivity
\cite{ferrer}. Finally, note that in the dense quark matter, 
a superstrong magnetic field is originated due to the presence of the
gluon
condensate \cite{iwazaki}. In conclusion, we see that there exist
several new physical effects that are intrinsically connected with
the gluon condensate (external chromomagnetic fields).
 
In the present paper, we study the influence of an external
chromomagnetic field on the pion condensation phenomenon. This phase
transition can 
also occur in the dense baryon matter, although isotopic
asymmetry, with different densities of up and down quarks, is
needed for this process to take place. This type of quark matter was
already
investigated  in the framework of NJL type of models, both with and
without pion condensation
\cite{bub,bed,barducci,zhuang,dekgk}. The main purpose of the present
paper is to investigate, in the framework of a NJL model, the
behavior of the quark and pion condensation in the presence of an 
external chromomagnetic field, modelling the gluon
condensate. In particular, we will show that, even for a weak coupling
of quarks, the pion condensation effect is 
induced by an external chromomagnetic field  and is related
to an effective dimensional reduction. 
The latter effect leads to a nonanalytic logarithmic dependence of
the quark and pion condensates on the field strength in the strong
field limit. 

\section{ Quark and pion condensates in external fields}

For a system, composed of two flavored quarks, there exists the relation 
$n_Q=n_{I_3}+n_B/2$ between the electric charge density $n_Q$, the
baryon charge density $n_B$ and the density
$n_{I_3}=\frac 12\bar q\tau_3q$ of the third isospin component
$I_3=\tau_3/2$. 
Since these quantities are linearly dependent, we will study in the
following the isotopic asymmetry (which means
that different species of quarks have different densities)
in dense quark matter
with a nonzero baryon chemical potential $\mu_B$.
In this case, the chemical potentials
$\mu_B$ and $\mu_I$ (where $\mu_I$ is the isospin chemical potential)
are independent parameters. Note that in another
possible case, i.e., $\mu_B\ne 0$, $\mu_Q\ne 0$
(the last quantity is the electric charge chemical potential),
they are no more independent, but related through the electric
charge neutrality condition $n_Q=0$, if one assumes that quarks are in a
weak-equilibrium with electrons, and the whole system is electrically
neutral. We restrict ourselves here to the
first possibility, i.e. suppose that $\mu_B$ and $\mu_I$ are
independent quantities. In our investigations we also suppose that
$\mu_B$ is a rather small quantity. In this case, as well as at
sufficiently high flavor asymmetry, the  color superconductivity
effects may
be neglected in favour of the normal quark matter or pion
superfluidity effects (see the recent discussions in
\cite{bed,zhuang}). 

\subsection{ General definitions}

Let us consider a NJL model of flavored and colored quarks $q_{i\alpha}\,
(i=1,\dots,N_f,\,\alpha=1,\dots,N_c)$ with $N_f=2, N_c=3$ as numbers
of flavors and colors, respectively
(for convenience, corresponding indices are
sometimes suppressed in what follows),
moving in an external chromomagnetic field.
The underlying quark Lagrangian
is chosen to contain four-quark interaction terms 
responsible for spontaneous breaking of chiral 
and flavor symmetries. Hence, two types of condensates might 
characterize the
ground state of the model: the quark
condensate $\vev{\bar qq}$ (spontaneous breaking of chiral
symmetry), and the pion condensate $\vev{\bar q \gamma_5\tau_1 q}$
(spontaneous breaking of parity and isotopic symmetry). In particular,
we consider a Lagrangian which describes dense quark matter
with an isotopical asymmetry, and we neglect diquark interactions
and
hence the possible formation
of a diquark condensate. Upon performing
the usual
bosonization procedure \cite{EbPerv}, \cite {7}
and introducing meson  fields $\sigma,\,\vec{\pi}$, the four-quark
terms are replaced by 
Yukawa interactions of quarks with these fields, and the Lagrangian
takes the following form
(our notations refer to four--dimensional Euclidean space with
$it=x_4$~\footnote{
We consider $\gamma-$matrices in the $4-$dimensional Euclidean space
with the metric tensor $g_{\mu\nu}=\diag (-1, -1, -1, -1)$, and the
relation between the Euclidean and Minkowski time
$x_{(E)}^0=ix_{(M)}^0
$: $ \gamma_{(E)}^0=i\gamma_{(M)}^0, \,
\gamma_{(E)}^k=\gamma_{(M)}^k.$ In what follows we denote the
Euclidean Dirac matrices as $\gamma_{\mu}$, suppressing the subscript
$(E).$ They have the following basic properties $\gamma_{\mu}^+= -
\gamma_{\mu},\,\{ \gamma_{\mu},\gamma_{\nu}\}=- 2\delta_{\mu\nu}.$
The charge conjugation operation for Dirac spinors is defined as $
\psi_c(x)\quad =\quad C\left(\psi(x)^+\right)^t $ with $
C\gamma_{\mu}^tC^{-1}=-\gamma_{\mu}.$ We choose the standard
representation for the Dirac matrices (see \cite{Rho}). The
$\gamma_5$
has the following properties:
$\{\gamma^{\mu},\gamma_5\}=0,\quad\gamma_5^+=\gamma_5^t=\gamma_5.$
Hence, one finds for the charge-conjugation matrix:
$C=\gamma^0\gamma^2,\quad C^+=C^{-1}=C^t=-C.$}):
\begin{eqnarray}
{\cal L} &=&-\bar q(i\gamma_\nu\nabla_\nu
+i\mu\gamma_0+\sigma+m +i\gamma^5\vec
\tau\vec\pi+i\mu'\tau_3\gamma^0)q-\frac{1}{4G}(\sigma^2+\vec \pi^2).
\label{1}
\end{eqnarray}
Here $\mu=(\mu_u  + \mu_d)/2$ is the chemical
potential averaged over flavors 
\footnote{It is equal to one third of the baryon chemical potential:
$\mu=\mu_B/3$.
Since the generator $I_3$ of the third component of isospin is equal
to $\tau_3/2$, the quantity $\mu'$ in (\ref{1}) is half 
the isospin chemical potential, $\mu'=\mu_I/2$.}, $\mu'=(\mu_u
-\mu_d)/2$ is their difference, and $G$ is the
(positive) four-quark coupling constant.
Furthermore,
$\nabla _\mu
=\partial_{\mu}-igA_{\mu}^a\lambda_a/2$ is the covariant
derivative of quark fields in the background field $F_{\mu
\nu }^a=\partial _\mu A_\nu ^a-\partial _\nu A_\mu ^a+gf_{abc}A_\mu
^bA_\nu ^c$ determined by the potentials $A_\mu ^a\left(
a=1,...,8\right)$, and
$\lambda_a/2$ are the generators of the
color $SU_c(3)$ group. Finally, $\vec \tau\equiv (\tau^{1},
\tau^{2},\tau^3)$ are Pauli
matrices in the flavor space.

Evidently, the Lagrangian (\ref{1}) with nonvanishing current quark
masses, $m\ne 0$, is invariant under the baryon
$ U_B(1)$ symmetry and the parity transformation P. Moreover,
without the $\mu'$-term, it
is also invariant under the isospin $SU_I(2)$ group which is
reduced to
$U_{I_3}(1)$ at $\mu'\ne 0$. At $m=0$ and $\mu'\ne 0$ the
symmetry group of the initial model is
$U_B(1)\times U_{I_3L}(1)\times U_{I_3R}(1)\times P$ (here the
subscripts $L,R$ mean that the corresponding group acts only on
the left, right handed spinors, respectively). It is very convenient
to present
this symmetry as $U_B(1)\times U_{I_3}(1)\times U_{AI_3}(1)\times
P$, where $U_{I_3}(1)$ is the isospin  subgroup and
$ U_{AI_3}(1)$ is the axial isospin subgroup. Quarks are
tranformed
under 
these 
groups in the following ways  $q\to\exp
(i\alpha\tau_3) q$ and $q\to\exp (i\alpha\gamma^5\tau_3) q$,
respectively. In the
case of 
$m= 0$
the
phase structure of the model (\ref{1}) is defined by the competition
of only two condensates. One of them is the quark condensate
$\vev{\bar qq}$, and the other is the pion condensate $\vev{\bar q
\gamma_5\tau_1 q}$. If the ground state of the model is characterized
by $\vev{\bar qq}\ne 0$ and $\vev{\bar q \gamma_5\tau_1 q}= 0$, then
the axial symmetry $ U_{AI_3}(1)$
of the model is spontaneously broken down, but isospin symmetry $
U_{I_3}(1)$ and parity P remain intact. However, if $\vev{\bar q
\gamma_5\tau_1 q}\ne 0$ and $\vev{\bar qq}= 0$,
then only parity P and the isospin symmetry are spontaneously
broken
down and the pion condensed phase is realized in the model. As a
consequence,  in the last case we have $\vev{\bar q \tau_3 q}\ne 0$,
i.e., a nonzero difference  in densities of up and down quarks arises
(isotopic asymmetry of the ground state). Generally speaking, the
inverse is not true, i.e., if there is an isotopic asymmetry of quark
matter, this does not necessarily mean that pion condensation
phenomenon does occur. 

In order to investigate the possible
generation of quark and pion condensates in the framework of the
initial model  
(\ref{1}), let us introduce the partition function $Z$ of the
system
\begin{equation}
Z =\exp W_{E}\,\,=\,\,\int dqd\bar{q}d\sigma d\pi_i  \exp
\left[\int d^4x{\cal L}\right]
\label{2}
\end{equation}
with $W_{E}$ being the Euclidean effective action.
Here, the meson fields can be decomposed as follows
\[
\sigma=\sigma^{(0)}+\delta \sigma,\,\,\,\vec \pi=\vec
\pi^{(0)}+\delta \vec \pi,
\]
where $\sigma^{(0)}$ and $\vec \pi^{(0)}$ are the condensate
fields, which  are
determined by the minimum of the effective action,
\[
{\delta\log Z\over \delta\sigma^{(0)}}=-\vev{\bar q q+{1\over
  2G}(\sigma^{(0)}+\delta \sigma)}=0,
\]
\[
{\delta\log Z\over \delta\vec\pi^{(0)}}=-\vev{\bar q i\gamma_5\vec
\tau 
q+{1\over 2G}(\vec\pi^{(0)}+\delta \vec\pi)}=0.
\]
Notice that the above expectation values are expressed through
functional integrals over the quark and meson  fields, where the
latter are calculated  
by using the saddle point approximation
(in what follows, we shall denote these mean 
values again by $\sigma$ and $\vec\pi$, respectively). Thus,
instead of
Eq. (\ref{2}),  we shall deal with the following functional
integral over fermion fields:
\begin{equation}
Z =\exp W_{E}\,\,=\,\,\int dqd\bar{q} \exp
\left[\int d^4x{\cal L}\right],
\label{2_}
\end{equation}
where now  $\sigma$ and $\vec\pi$ in the Lagrangian (\ref {1}) are
understood as constant
condensates. 
 In the mean field approximation, where field fluctuations
$\delta \sigma$ and $\delta \vec \pi$ are neglected, they are given by
 the following gap equations
\begin{eqnarray}
-\frac {1}{2G}\sigma=\vev{\bar{q} q},\,\,\,\,
-\frac {1}{2G}\vec\pi=\vev{i\bar{q}\vec \tau\gamma_5 q}.
\label{delta1}
\end{eqnarray}
Assume that the only
nonvanishing components of the background gauge field potential are
for $a=1,2,3$, while others are equal to zero,
\footnote{Of course, if the external constant homogeneous gauge field
fills the whole space, then the color as well as the rotational
symmetries of the system are broken. This might be considered as an
artifact, since in reality there exist both color and rotational
invariance (we ignore the color superconductivity effects). In our
case, in order to deal with a physically acceptable ground state,
i.e. without color symmetry breaking, one can interpret it as a
space split
into an infinite number of domains with macroscopic extension. Inside
each such domain there exist a homogeneous background 
chromomagnetic field, which generates a nonzero gluon condensate
$\vev{F^a_{\mu\nu}F^{a\mu\nu}}$. Moreover, the direction of the gauge
field varies from domain to domain, so the averaging over all domains
results in a zero background chromomagnetic field. Therefore, color
as
well as rotational symmetries are not broken. Strictly speaking, our
following calculations refer to some given macroscopic domain.  The
obtained results turn out to depend on color and rotational invariant
quantities only, and are independent of the particular domain.}
\[A_{\mu}^a\ne
0,\,{\rm when}\,\,a=1,2,3,\,\,{\rm and}\,\, A_{\mu}^a=0,\,{\rm
when}\,\,a=4,\dots,8.\] 
This implies
that only quarks of two colors $\alpha =1,2$ do interact with the
background field $A_{\mu}^a$.
As a result,
the integration over quark degrees of freedom in the partition
function (\ref{2_}) is greatly simplified, and we have 
\begin{equation}
Z=\Det_{(1)}(i\gamma\partial+{\cal M})
\cdot\Det_{(2)}(i\gamma\nabla+{\cal M}),
\label{intermediate}
\end{equation}
where ${\cal M}=m+\sigma
+i\gamma^5\vec\pi\vec\tau+i\mu\gamma_0+i\mu'\tau_3\gamma^0$, and
indices (1) and (2) mean that determinants are calculated in the
one-dimensional (with color $\alpha=3$) and in the
two-dimensional (with colors $\alpha=1,2$) subspaces of the color
group, respectively. 
Assume now that the background field is constant and homogeneous,
$F_{\mu\nu}^a=const$. Then the Dirac equations 
\[
\left(i\gamma \partial + {\cal M}\right)\psi=0 ,
~~~~~~~~\left(i\gamma \nabla + {\cal M}\right)\psi=0 
\] 
have stationary
solutions with the energy spectrum $\varepsilon$ 
for quarks of flavor $i$ and color $\alpha=1,2,3$ with quantum
numbers $k$
moving  in the constant background field $F_{\mu\nu}^{a} (a=1,2,3)$. 
In this case, we arrive at the following Euclidean effective
action: \begin{equation} 
W_E=\tau\int \frac{dp_4}{2\pi
}\sum_{\lambda,\alpha,k,\kappa}\log
\left[p_4^2+(\varepsilon-\kappa\mu)^2\right] - (\tau
L^3){\sigma^2+\vec \pi^2\over 4G}.
\label{3}
\end{equation}
Here,  $\tau$ is the imaginary time interval, the sum is over the
signs $\lambda=\pm 1$ of the 
chemical potential $\mu'$, the signs $\kappa=\pm1$ of the
chemical potential $\mu$, corresponding
to charge conjugate contributions of 
quarks,  color indices $\alpha=1,2,3,$ and also over quantum numbers
$k$ of quarks, with $\alpha=3$ for free quarks 
and the spectrum
\begin{equation}
\eps=\varepsilon_{\vec p, \lambda}=\sqrt{\Big
(\sqrt{(\sigma+m)^2+\vec p^2
+\pi_3^2}+\lambda\mu'\Big )^2+\pi_1^2+\pi_2^2},\quad (\lambda=\pm 1),
\end{equation}
and with $\alpha=1,2$ for quarks with the spectrum
\begin{eqnarray}
\eps=\varepsilon_{k, \lambda}=\sqrt{\Big
(\sqrt{(\sigma+m)^2+{\Pi}_k^2
+\pi_3^2}+\lambda\mu'\Big )^2+\pi_1^2+\pi_2^2}, \quad (\lambda=\pm 1),
\label{45}
\end{eqnarray}
 moving in the background color field
$F_{\mu\nu}^a\,(a=1,2,3)$. In the above formula, ${\Pi}_k^2$ stands
for
 the eigenvalues of the squared Dirac operator $-(\vec\gamma\vec
 \nabla)^2$
 with $\vec \nabla=\vec \partial - ig\vec A^a \lambda_a/2$.

In the case of finite temperature $T= 1/
\beta >0$, the thermodynamic potential $\Omega=-W_E/(\beta L^{3})$
is obtained after substituting $p_4\rightarrow \frac{2\pi
}\beta (l+\frac 12), l=0,\pm 1,\pm 2,...$,
\begin{eqnarray}
\Omega&=&-\frac 1{\beta L^{3}}\sum_{\lambda,\kappa}\sum_{k,\alpha}
\sum^{l=+\infty}_{l=-
\infty}\log\left[\left(\frac{2\pi(l+1/2)}{\beta
}\right)^2+(\eps-\kappa\mu)^2\right]+\frac{\sigma^2+\vec{\pi}^2}{4G}
+ C, \label{450}
\end{eqnarray}
where we introduced a subtraction constant $C$ in such a way that at $
 \sigma=\vec{\pi}=0$ we have 
 $ \Omega=0 $.
Next, let us consider the proper time representation
\begin{equation}
\Omega = \frac 1{\beta
  L^{3}}\sum_{\lambda,\kappa}\sum_{k,\alpha}\sum^{l=+\infty}_{l=-
  \infty} 
\ints^{\infty}_{1/\Lambda_s^2}\frac{ds}s
\exp\biggl[-s\left(\frac{2\pi(l+1/2)}{\beta}\right)^2
  -s(\eps-\kappa\mu)^2\biggr]+\frac{\sigma^2+\vec{\pi}^2}{4G} + C, 
\label{mu}
\end{equation}
where $\Lambda_s $ is an ultraviolet cutoff ($\Lambda_s \gg
\sigma,\,|\vec\pi|$). 
The 
temperature dependent contribution can be further transformed with 
the help of the formula 
\begin{equation} 
\label{eq19} 
\sum _{l=-\infty }^{+\infty }\exp [-s(2\pi l/\beta +x)^2]=\frac
{\beta } 
{2 \sqrt{\pi s}}[1+2\sum _{l=1}^{\infty }\exp (-\frac{\beta
^2l^2}{4s}) 
\cos (x\beta l)],
\end{equation}
where in our case $x={\pi\over\beta}$. Then,  calculating the
quark condensate
\begin{equation} \langle \bar{q}q\rangle =
\frac{\ds\int\,d\bar qdq ~~ \bar{q}q \exp\left[\ds\int\,d^4x {\cal
L} \right]} {\ds\int\,d\bar{q}dq \exp \left[\ds\int\,d^4x {\cal
L} \right]},
\label{5}
\end{equation}
and combining the result of (\ref{5}) and (\ref{delta1}), we obtain 
the following gap equation:
\begin{equation} \frac
{\sigma}{G} = 
\frac{\sigma+m}{L^{3}\sqrt{\pi}}\sum_{\lambda,\kappa}\sum_{k,\alpha}
\ints^{\infty}_{1/\Lambda_s^2}
\frac{\ds
ds}{\sqrt{s}}\left[1+2\sum^{\infty}_{l=1}(-1)^l
\e^{\ts-\frac{\beta^2l^2}{4s}}\right]\e^{\ts-s(\eps-\kappa\mu)^2}
\frac {(\eps -
  \kappa\mu)\sqrt{\varepsilon^2-\pi_1^2-\pi_2^2}}{\eps(\sqrt{
  \varepsilon^2-\pi_1^2-\pi_2^2}-\lambda\mu')}. 
\label{6}
\end{equation}
Here, the first term in the square brackets describes the $T=0$
contribution (it corresponds to the result of integration over $p_4$
in the initial equation (\ref{3})), while the second term is the
finite temperature
contribution, $T\ne 0$.
The $\vec\pi$-condensate 
can be obtained in a similar way. For $\pi_{3}$ we have
\begin{equation}
\frac {\pi_{3}}{G}\,=\,
\frac{\pi_{3}}{L^{3}\sqrt{\pi}}\sum_{\lambda,\kappa}\sum_{k,\alpha}
\ints^{\infty}_{1/\Lambda_s^2}
\frac{\ds
ds}{\sqrt{s}}\left[1+2\sum^{\infty}_{l=1}(-1)^l
\e^{\ts-\frac{\beta^2l^2}{4s}}\right]\e^{\ts-s(\eps-\kappa\mu)^2}
\frac {(\eps -
  \kappa\mu)\sqrt{\varepsilon^2-\pi_1^2-\pi_2^2}}{\eps(\sqrt{
  \varepsilon^2-\pi_1^2-\pi_2^2}-\lambda\mu')}.
\label{qq}  
\end{equation}
By comparing this expression with that for $\sigma$, we conclude that
for nonvanishing quark mass $m\ne 0$ this condensate vanishes,
$\pi_3=0$. For 
$\pi_1$ (putting $\pi_2=0$ by consideration of the symmetry of the
problem) we have 
\begin{equation}
\frac
{\pi_1}{G}\,=\,\frac{\pi_1}{L^{3}\sqrt{\pi}}\sum_{\lambda,\kappa}\sum
_{k,\alpha}\ints^{\infty}_{1/\Lambda_s^2}
\frac{\ds
ds}{\sqrt{s}}\left[1+2\sum^{\infty}_{l=1}(-1)^l
\e^{\ts-\frac{\beta^2l^2}{4s}}\right]\frac {(\eps -
  \kappa\mu)}{\eps}\e^{\ts-s(\eps-\kappa\mu)^2}.\label{qqq}
\end{equation}
It follows from (\ref{6}) and (\ref{qqq}) that at $m=0$ and $\mu'\ne
0$ there exist only two different 
solutions of this system of gap equations 
(except for the trivial one with $\sigma = \pi_1 =0$), i.e., a)
$\sigma =0, \pi_1 \ne 0$ and b) $\sigma \ne 0, \pi_1=0$. Thus we
have to find out which of these 
solutions provide the global minimum of the 
thermodynamic potential (\ref{450}) with $\mu'\ne 0$. 

In the  limit of a vanishing external
field ($F_{\mu \nu }^a=0$), with $\pi_3=0,\,\,\pi_2=0$, we have for
the
particle spectrum
\begin{equation}
\varepsilon_{\vec p, \lambda}=\sqrt{\Big (\sqrt{(\sigma+m)^2+\vec p^2
}+\lambda\mu'\Big )^2+\pi_1^2},
\label{free}
\end{equation}
and for the sum over quantum states
$$
\frac {1}{
L^3}\sum_{k,\alpha,\kappa,\lambda}\,=\,12\sum_{\lambda}\int\frac{d^3p
}{(2\pi)^3}
$$
for 3 color states ($\alpha=1,2,3$), 2 spin states, and  2 values of
$\kappa=\pm 1$.
Considering now the case of vanishing temperature $T=0$, we shall
omit the cutoff in the lower
limit, $1/\Lambda_s^2\to 0$, and replace it by the corresponding
cutoff in the
momentum integration $\vec {p}^2\le \Lambda_p^2$. It can be
easily seen that
the two cutoff parameters $\Lambda_p$ and $\Lambda_s$ are related as
$\Lambda_s^2=2\Lambda_p^2$. Indeed, integration in (\ref{3}) gives:
$$
-\frac{1}{(2\pi)^4}\int dp_4\ints_{\vec p^2<\Lambda_p^2}d^3p\log(\varepsilon^2+p_4^2)=
-\frac{1}{8\pi^2}\Lambda_p^4 + O(\Lambda_p^2)
$$
with the momentum cutoff $\vec {p}^2\le \Lambda_p^2$, and integration
in (\ref{mu}) gives
$$
\frac{1}{(2\pi)^4}\int
d^4p\ints_{1/\Lambda_s^2}^{\infty}\frac{dz}{z}\exp(-iz\eps^2-iz
p_4^2)=-\frac{1}{32\pi^2}\Lambda_s^4+ O(\Lambda_s^2)
$$
with the proper time cutoff $z\ge \frac{1}{\Lambda_s^2}$. In what
follows, we shall denote the $\Lambda_p^2$ cutoff by $\Lambda^2$,
and we 
shall use it throughout, substituting $\Lambda_s^2=2\Lambda^2$. Now,
we can perform the proper-time integration 
in (\ref{qqq}) with the help of the formula
\begin{equation}
\sum_{\kappa=\pm 1}\ints^{\infty}_{0}\frac{\ds ds}{\sqrt{s}}\frac
{(\eps -
\kappa\mu)}{\eps}\e^{\ts-s(\eps-\kappa\mu)^2}=\left(\frac{\eps-\mu}{|
\eps-\mu|}+1\right)\frac{\sqrt{
\pi}}{2\eps}=\sqrt{\pi}\frac{\theta(\eps-\mu)}{\eps}.
\end{equation}
As a result, one obtains for the quark and pion condensates in the
vanishing
color field background the formulas coinciding with the result of
\cite{dekgk},
\begin{eqnarray}
&&\frac{\sigma}{2G}=6(\sigma+m)\sum_{\lambda}\ints_{\vec {p}^2\le
\Lambda^2}\frac{d^3p}{(2\pi)^3}
\frac{\theta (\varepsilon-\mu)}{\varepsilon}
\frac{\sqrt{(\sigma+m)^2+\vec p^2}
+\lambda\mu'}{\sqrt{(\sigma+m)^2+\vec p^2}},
\label{50}\\
&&\frac{\pi_1}{2G}=6\pi_1\sum_{\lambda}\ints_{\vec {p}^2\le
\Lambda^2}\frac{d^3p}{(2\pi)^3}
\frac{\theta (\varepsilon-\mu)}{\varepsilon}.
\label{51}
\end{eqnarray}
In what follows, we shall analyze 
the special case of a constant Abelian chromomagnetic field
\begin{eqnarray}
A_\mu ^a&=&\delta^a_3\delta_{\mu2}x_1H.
\label{13}
\end{eqnarray}
The  spectrum $\Pi_k^2$\,\, of the Dirac operator\,\,  $-(\vec
\gamma\vec \nabla)^2$ has then six branches, two of which correspond
to quarks that
do not interact with the chromomagnetic field ($\alpha=3$)
\begin{eqnarray}
{\Pi}_{1,2}^2={\vec p}^2,
\label{9}
\end{eqnarray}
and the other four correspond to two color degrees of
freedom of quarks with ``charges'' $\pm g/2$
interacting with the external field. The ${\Pi}_k^2$ spectrum of
quarks is now given by ($\alpha=1,2$)
\begin{eqnarray}
{\Pi}_{3,4,5,6}^2\,=\,gH(n+\frac12+
\frac{\zeta}2)+p^2_3,
\label{14}
\end{eqnarray}
where $\zeta =\pm 1$ is the spin projection on the external field
direction, $p_3$ is the longitudinal component of the quark momentum
($-\infty <p_3<\infty $),
\begin{eqnarray}
p_{\perp}^2=gH(n+\frac12)
\label{15}
\end{eqnarray}
is the transversal component squared of the quark momentum, and
$n=0,1,2,...$
is the Landau quantum number.

The form of the spectrum
is essential for the quark and pion condensate formation. Using the
above
expressions for energy spectra, we shall next study the quark and
pion
condensates in the strong field limit. 

\subsection{ Asymptotic estimates for strong fields}

In this section, we consider the special cases of the above
configurations of background fields in the strong field
limit. Our goal is
here  to demonstrate that the field is a catalyzing agent for
dynamical symmetry breaking, leading to a possible creation of
corresponding condensates.  The 
external 
fields are assumed
to be strong as compared to the values of 
quark $\vev{\bar qq}$ and pion $\vev{\bar q \gamma_5\tau_1 q}$
condensates that may be
rather small. In this sense, even the expected values of fields
simulating the presence of a gluon condensate, which we take to be of
the order of  $gH=0.4-0.6\,\, {\rm
GeV}^2$, may be considered as strong 
(these values of the external chromomagnetic field $gH$ correspond to
the QCD gluon condensate at zero temperature and zero baryon density
\cite{nar}).
As for the other parameters, we may take their values as in
\cite{dekgk,zhuang}, i.e., $\Lambda=0.65$ GeV,
$G=5.01$ GeV$^{-2}$. 

Consider now the special choice of parameters:
$\mu\,=\pi_2\,=\,\pi_3\,=\,m\,=\,0.$  In this simple massless case,
though with $\pi_1\,\ne\, 0,\,\,\sigma\,\ne\, 0,\,\,\mu'\,\ne\, 0, $
the thermodynamic potential (\ref{mu}) takes the form  
\begin{equation}
\Omega = \frac 1{2\sqrt {\pi}
  L^{3}}\sum_{\lambda,\kappa}\sum_{k,\alpha}
\ints^{\infty}_{1\over 2\Lambda^2}\frac{ds}{s^{\frac 3 2}}
\left[1+2\sum^{\infty}_{l=1}(-1)^l
\e^{\ts-\frac{\beta^2l^2}{4s}}\right]\e^{\ts
  -s\eps^2}+\frac{\sigma^2+\pi_1^2}{4G} + C.  
\label{muu}
\end{equation}

Next, consider the zero temperature case, $T=0$. Then the second
term in the square bracket in the above expression vanishes and we are
left
with the first term equal to unity. The above equation describes
the effective potential  $V_{\rm eff}=\Omega|_{T=0}$
\begin{equation}
V_{\rm eff}(\sigma, \pi_1) = \frac 1{2\sqrt {\pi}
  }\ints^{\infty}_{1\over 2\Lambda^2}\frac{ds}{s^{\frac 3 2}}
\Big[
4\sum_{\lambda}\int\frac{d^3p}{(2\pi)^3}
\e^{\ts-s\varepsilon_{\vec p,
  \lambda}^2}\,+\,\frac {1}
  {L^{3}}\sum_{\kappa}\sum_{k,\lambda,\alpha=1,2}\e^{\ts-s\varepsilon
  _{k,\lambda}^2}\Big]+\frac{\sigma^2+\pi_1^2}{4G} + C.
\label{muuu}
\end{equation}
The first term in the square bracket stands for quarks with
$\alpha=3$, when we have
\begin{equation}
\varepsilon_{\vec p, \lambda}^2=
(\sqrt{\sigma^2+{\vec p}^2}+\lambda\mu' )^2+\pi_1^2,
\end{equation}
and 4 stands for two values of $\kappa=\pm 1$ and two values of spin
projection $\zeta=\pm 1$. The second term stands for quarks with
$\alpha=1,2$, where we have 
\begin{equation}
\varepsilon_{k,\lambda}^2=(\sqrt{\sigma^2+\Pi_k^2}+\lambda\mu'
)^2+\pi_1^2.
\end{equation}

Let us first consider the symmetric case, when the chemical potentials
for the $u$-quark and the $d$-quark are equal,
$\mu'=(\mu_u-\mu_d)/2=0$. Then, the energy of the free quark with
$\alpha=3$ becomes 
\[\varepsilon_{\vec p, \lambda}^2={\vec
  p}^2+\sigma^2+\pi_1^2={\vec p}^2+\sigma_1^2,\] 
where $\sigma_1^2=\sigma^2+\pi_1^2$. 

Consider now the case, when the Abelian-like background
field is strong enough, $gH=O(\Lambda^2)$, but 
$gH<2\Lambda^2$. 
The momentum squared $\Pi_k^2$ is given by (\ref{14}), and summation
over quark quantum numbers in the chromomagnetic field gives 
\begin{equation}
\frac {1}
{L^{3}}\sum_{\lambda}\sum_{k,\alpha}=2\frac{gH}{4\pi}\sum_{\lambda}
\sum_{n=0}
^{\infty}(2-\delta_{n0})\int\frac{dp_3}{2\pi}.   
\label{degenerate}
\end{equation}
Now, the  momentum integral in the first term in (\ref{muuu})
gives
\[
\int\frac{d^3p}{(2\pi)^3}\e^{\ts-s\varepsilon_{\vec p,
\lambda}^2}=\frac{\sqrt{\pi}}{8\pi^2}\frac{\e^{\ts-s\sigma_1^2}}{s^{3
/2}}.
\] 
In the second term in (\ref{muuu}) for quarks with $\alpha=1,2$, due
to
the strong field condition, we take only the contribution of the
term with $n=0$ in the sum over Landau quantum numbers, while
the contribution of large quantum numbers is described similar to the
contribution of the free
quark term, and thus we must take the free term three
times. As a result, after integration in the second term over the 
$p_3$-component of the quark momentum we have 
\[
V_{\rm
  eff}(\sigma_1)=\frac{1}{4\pi^2}\left[\ints_{1\over
  2\Lambda^2}^{\infty}ds\e^{\ts-
  s\sigma_1^2}\frac{3}{s^3}+
\ints_{1\over
gH}^{\infty}ds\e^{\ts-s\sigma_1^2}\frac{gH}{2s^2}\right]+
\frac{\sigma_1^2}{4G}+C.   
\]
The cutoff momentum $\Lambda$ as well as the background field $gH$
are large $\Lambda\gg\sigma_1,\,\,gH\gg\sigma_1^2$ and, hence we can
approximately integrate over $s$ in the first and the second terms
according to the following formulas:
\begin{equation}
\ints_{x_{\rm min}}^{\infty}\frac{dx}{x^3}\e^{\ts-x}={1\over2x_{\rm
    min}^2}-{1\over x_{\rm min}}-{1\over2}\log x_{\rm min}+{\rm
    const},\,\,\,\mbox{where}\,\,x_{\rm
    min}={\sigma_1^2\over 2\Lambda^2}\ll 1,
\label{lambda}
\end{equation}
\begin{equation}
\ints_{x_{\rm min}}^{\infty}\frac{dx}{x^2}\e^{\ts-x}={1\over x_{\rm
    min}}+\log x_{\rm min}+{\rm const}, \,\,\,\mbox{where}\,\, x_{\rm
    min}={\sigma_1^2\over gH}\ll 1.
\label{gH}
\end{equation}
As a result we find the effective potential
\begin{equation}
V_{\rm
  eff}
  (\sigma_1)=\frac{1}{4\pi^2}\Bigl\{3\sigma_1^4\left[{1\over2}\left({
 2 \Lambda^2\over\sigma_1^2}\right)^2
-{2\Lambda^2\over\sigma_1^2}+{1\over2}\log
  {2\Lambda^2\over\sigma_1^2}+C_1\right]+{gH\sigma_1^2\over2}\left({
  gH\over\sigma_1^2}+\log{\sigma_1^2\over
gH}+C_2\right) \Bigr\}+{\sigma_1^2\over 4G}+C,\label{kl}
\end{equation}
where $C_1,\,\,C_2$ are certain numerical constants. 
Now, let us find the minimum $\sigma_0$ of the thermodynamic
potential (\ref{kl}). Then, the minimum $\sigma_0$ is described by the
solution of the equation, where we neglect the terms that do not contain large
parameters $\Lambda^2$ or $gH\log(gH/\sigma_0^2)$,
\[
{\partial V_{\rm eff}\over\partial \sigma_1^2} |_{\sigma_1=\sigma_0}=
{1\over4\pi^2}\left[-6\Lambda^2-3\sigma_0^2\log
  {\sigma_0^2\over 2\Lambda^2}+{gH\over 2}\log{\sigma_0^2\over
  gH}\right]+{1\over 4G}=0.
\]
Within the approximation of strong background field
\begin{equation}
gH\log(gH/\sigma_0^2)\gg\sigma_0^2\log(2\Lambda^2/\sigma_0^2),
\label{strong}
\end{equation}
the above equation simplifies to the following form:
\begin{equation}
1={6G\Lambda^2\over \pi^2}+G{gH\over 2\pi^2}\log{gH\over\sigma_0^2}
\label{cutoff}
\end{equation}
with the solution
\begin{equation}
\sigma_0=\sqrt{\frac{gH}{2\pi}}\exp\left(-\frac{2\pi^2(1-\tilde
    g)}{GgH}\right), 
\label{piv1}
\end{equation}
where the effective coupling constant, 
\begin{equation}
\tilde g =\frac {6G\Lambda^2}{\pi^2},
\label{tg}
\end{equation}
can be arbitrary small $\tilde g <1$, contrary to the zero magnetic
field case $H=0$, when it should be $\tilde g >1$  in order that the
nontrivial solution for $\sigma_0$ exists.
This is a very interesting result. Indeed, we see that at $\mu'=0$ and
arbitrary
small attraction between quarks ($\tilde g <1$) the external
chromomagnetic field catalyzes the dynamical chiral symmetry breaking
and quarks acquire a nonzero mass.
which is equal to $\sigma_0$ (\ref{piv1}). 
In contrast, if $H=0$ and $\tilde g <1$, the quarks remain massless
and symmetry is unbroken.
After removing the cutoff with the use of (\ref{cutoff}), the
effective potential (\ref{kl}) takes the form
\begin{equation}
V_{\rm eff}(\sigma_1)={gH\sigma_1^2\over 8\pi^2}\left(\log
{\sigma_1^2\over
  \sigma_0^2}-1\right),
\label{eff}
\end{equation} 
which is just the effective potential of the two-dimensional
Gross-Neveu model (see, e.g., \cite{bieten}). 
This result shows that the chromomagnetic catalysis effect is
accompanied by the effective reduction of the space-time
dimensionality.   

Next, consider the case of flavor asymmetric quark matter with
$\mu'\ne 0$. 
For simplicity, $\mu$ is taken to be zero as before.
In this case (recall that $m=0$), there are two nontrivial
solutions of the gap equations (\ref{6}) and (\ref{qqq}) (see the
remark after (\ref{qqq})), which are the points on the $\sigma$- or
$\pi_1$-axes. So, in order to find the global minimum point, it is
sufficient (and very convenient) to reduce the investigation of
the effective potential $V_{\rm eff}(\sigma, \pi_1)$ (\ref{muuu}) as
a function of two variables 
$\sigma$ and $\pi_1$ to two particular cases. First, we
shall study $V_{\rm eff}$ as a function of $\pi_1$ only with  
$\sigma=0,\,\,\pi_1\ne 0$, then as a function of 
$\sigma$ with $\sigma \ne 0,\,\,\pi_1=0$.
When comparing these two particular cases, one can obtain the
genuine global minimum point of the effective potential $V_{\rm
eff}(\sigma, \pi_1)$.
To simplify our calculation, we assume that $\mu'$ and $gH$ are
large, $\mu'^2=O(\Lambda^2)$ and $gH=O(\Lambda^2)$.

{\bf a) The case $\sigma=0,\,\,\pi_1\ne 0$.} 

In this case the first term in (\ref{muuu}),
\begin{equation}  
V_{\rm eff}^{(1)}(\pi_1) = \frac 1{2\sqrt {\pi}
  }\ints^{\infty}_{1\over 2\Lambda^2}\frac{ds}{s^{\frac 3 2}}
\Big[
4\sum_{\lambda}\int\frac{d^3p}{(2\pi)^3}
\e^{\ts-s\varepsilon_{\vec p,
  \lambda}^2}\,\Big]
\end{equation}
can be expressed through the following approximation of the integral
(with $s\mu'^2\gg 1$):
\begin{equation}
\sum_{\lambda}\int\frac{d^3p}{(2\pi)^3}
\e^{\ts-s\varepsilon_{\vec p,
    \lambda}^2}={\sqrt{\pi}\e^{\ts-s\pi_1^2}\over4\pi^2}\left(2\mu'^2
    +{1\over 2s}\right){1\over\sqrt{s}}.
\label{approx}
\end{equation}
Integration over $s$ should be approximately performed separately for
the first and second terms in (\ref{approx}) with different lower
limits
\[
V_{\rm eff}^{(1)}(\pi_1)={1\over
4
\pi^2}\left(4\ints^{\infty}_{1\over\mu'^2}\frac{ds\mu'^2}{s^{2}}\e^{\
ts-s
\pi_1^2}+3\ints^{\infty}_{1\over 2\Lambda^2}\frac{ds}{s^{3}}\e^{\ts-s
\pi_1^2}\right).
\] 
Here, we also added the contribution of quarks with $\alpha=1,2$ with
high Landau quantum numbers $n\gg 1$. 

In the second term in (\ref{muuu}),
\[
V_{\rm eff}^{(2)}(\pi_1) = \frac 1{2\sqrt {\pi}
  }\ints^{\infty}_{1/\Lambda_s^2}\frac{ds}{s^{\frac 3 2}}
\frac {1}
{L^{3}}\sum_{\kappa}\sum_{k,\lambda,\alpha=1,2}\e^{\ts-s
\varepsilon_{k,\lambda}^2},
\]
with  $gH=O(\Lambda^2)$ and $\mu'=O(\Lambda)$, but $gH < \mu'^2 <
2\Lambda^2$, we should
estimate the contribution of the two main terms, $n=0$ and $n=1$, in
the sum over $n$.
 The $n=0$
contribution is expressed through the integral
\begin{equation}
\sum_{\lambda}\int\frac{dp_3}{(2\pi)}\exp[-s(\pi_1^2+(|p_3|+\lambda
\mu')^2)]={1\over\pi}\e^{\ts-s\pi_1^2}\sqrt{\frac{\pi}{s}},
\label{easy}
\end{equation}
and summation over the remaining quantum numbers is made according to
(\ref{degenerate}) with $n=0$. For the $n=1$ term, we have a 
new expression with integration of the following exponential term:
\begin{equation}
\sum_{\lambda}\int\frac{dp_3}{(2\pi)}
\exp[-s(\pi_1^2+(\sqrt{gH+p_3^2}+\lambda\mu')^2)].
\label{expo}
\end{equation}
In order to estimate the $n=1$ contribution, we write $p_3=\bar p_3
+\psi$, where $\psi$ is a small deviation of $p_3$ from the minimum
of
the exponential quantity $\sqrt{\bar p_3^2+gH}=\mu'$ for
$\lambda=-1$. Then the integral in (\ref{expo}) for $n=1$ is
estimated as 
\begin{equation}
\sum_{\lambda}\int \frac{d\psi}{2\pi}\exp[-s \frac
  {\mu'^2-gH}{\mu'^2}\psi^2]\,=\,{1\over\pi}\sqrt{\frac{\pi}{s}}\,\,
  \frac{\mu'}{\sqrt{\mu'^2-gH}}.
\label{n1}
\end{equation}
Recall that this estimation is justified under the above assumption,
i.e.,  $gH=O(\Lambda^2)$ and $\mu'=O(\Lambda)$, but $gH < \mu'^2 <
2\Lambda^2$.
Finally,
summing up the $n=0$ and $n=1$ contributions (\ref{easy}) and 
(\ref{n1}), we find
\[
V_{\rm eff}^{(2)}(\pi_1)={gH\over 8\pi^2}\left(1+{2\mu'\over \sqrt
    {\mu'^2-gH}}\right)\ints^{\infty}_{1\over
    gH}\frac{ds}{s^{2}}\e^{-s
        \pi_1^2}.\] 
In the second term in the above bracket, we took into
    account that the state with $n=1$ is two-fold degenerate in the
    spin variable. Summing up all the contributions, and taking into
    account the integrals (\ref{lambda}) and (\ref{gH}), we have
\[
V_{\rm eff}(\pi_1)={1\over 4\pi^2}\Bigl\{{3\pi_1^4\over
2}\left[({2\Lambda^2\over
  \pi_1^2})^2-2{2\Lambda^2\over \pi_1^2}-\log {\pi_1^2\over
  2\Lambda^2}+C_1\right]
  +\left[4\mu'^4+{(gH)^2\over2}\left(1+{\mu'\over
  \sqrt{\mu'^2-gH}}\right)\right]+\]
\begin{equation}
  +4\mu'^2\pi_1^2\left(\log {\pi_1^2\over \mu'^2}+
  C_2\right)+{gH\pi_1^2\over2}\left(1+{\mu'\over\sqrt{\mu'^2-gH}}
  \right)\left(\log {\pi_1^2\over gH}+
  C_3\right)\Bigr\}+{\pi_1^2\over 4G}+ C.
\label{effpot}
\end{equation}
The minimum point $\pi_0$ of this function obeys the following
stationarity equation $\partial V_{\rm eff}/\partial
\pi_1^2|_{\pi_1=\pi_0}=0$, which gives
\begin{equation}
1={6G\Lambda^2\over \pi^2}-{4\mu'^2G\over \pi^2}\log {\pi_0^2\over
\mu'^2}-{gH\over
 2\pi^2}G\left(1+{\mu'\over\sqrt{\mu'^2-gH}}\right)\log {\pi_0^2\over
 gH},
\label{eqv}
\end{equation}
where the approximation of strong
background field,
$gH\log(gH/\pi_1^2)\gg\pi_1^2\log(2\Lambda^2/\pi_1^2),$ has been
used,
similar to (\ref{strong}).
After substituting the above equation into (\ref{effpot}), the
effective
potential takes the form
\begin{equation}
V_{\rm eff}(\pi_1^2)={\pi_1^2\over 4\pi^2}\left[4\mu'^2+{gH\over
  2}\left(1+{\mu'\over
  \sqrt{\mu'^2-gH}}\right)\right]\left(\log{\pi_1^2\over
  \pi_0^2}-1\right),
\label{effpoto}
\end{equation}
which again, like (\ref{eff}), resembles the effective potential in
the two-dimensional Gross
-- Neveu model \cite{bieten}. Note also that the minimum $\pi_0$ of
the function (\ref{effpot}), under our assumption of a strong color
background field,  exists even for weak coupling of quarks, i.e., at
$\tilde g <1$, while with zero background field, it can exist only if
$\tilde g >1$.
 Next, let us study the behaviour of the effective potential
 (\ref{muuu}) as a function of $\sigma$, when $\pi_1=0$. 

{\bf b) The case $\sigma\ne 0, \pi_1=0$.}

Here again we use the above assumption $gH=O(\Lambda^2)$ and
$\mu'=O(\Lambda)$, but $gH < \mu'^2 < 2\Lambda^2$.  
The term in the effective potential with $\alpha=3$ is determined by
the expression
\begin{equation}  
V_{\rm eff}^{(1)} = \frac 1{2\sqrt {\pi}
  }\ints^{\infty}_{1/\Lambda_s^2}\frac{ds}{s^{\frac 3 2}}
\Big[
4\sum_{\lambda}\int\frac{d^3p}{(2\pi)^3}
\e^{\ts-s\varepsilon_{\vec p,
  \lambda}^2}\,\Big],
\end{equation}
where $\varepsilon_{\vec p,
  \lambda}=\sqrt {\sigma^2+{\vec p}^2}+\lambda \mu'$. The main
  contribution to the integral over $p$, for large $\mu'$, is given
  by large $p$ near ${\overline p}=\sqrt{\mu'^2-\sigma^2}.$ Then the
  lower limit
  in the integral over $s$ should be taken as ${1\over
  \mu'^2-\sigma^2}$, and
  we obtain, with $p={\overline p}+\psi$,
\begin{equation} 
V_{\rm eff}^{(1)}={\bar p^2\over \pi^{5/2}}\ints_{1\over
  \mu'^2-\sigma^2}^{\infty}{ds\over s^{3/2}}\int d\psi \e^{\ts
  -s(\psi\bar
  p/\mu')^2}={\bar p \mu'\over\pi^2}\ints_{1\over
  \mu'^2-\sigma^2}^{\infty}{ds\over
  s^2}={1\over \pi^2}\mu'(\mu'^2-\sigma^2)^{3\over 2}.
\end{equation}

Then, the first term, where the contribution   of  quarks with $\alpha
  =1,2,3$ in the region of very large values of $s \gg {1\over \mu'^2},
  {1\over gH}$ is also added, is estimated as
\begin{equation}
V_{\rm eff 1}(\sigma)= {1\over \pi^2}(\mu'^2-\sigma
^2)^{3\over2}\mu'+{3\sigma^4\over 8\pi^2}\left[\left({2\Lambda^2\over
    \sigma^2}\right)^2-4{\Lambda^2\over \sigma^2}-\log {\sigma^2\over
     2 \Lambda^2}+C_1\right].
\label{eff1}
\end{equation}
Consider now the contribution of the term with $\alpha =1,2$ in the
region of comparatively small values of $s$ close to the threshold $1/gH$,
\begin{equation}
V_{\rm eff 2}={1\over 4
  \sqrt{\pi}}\ints^{\infty}_{1/gH}\frac{ds}{s^{\frac 3
  2}}\sum_{\lambda}{2gH\over 4\pi}\sum _n
  (2-\delta_{n0})\int{dp_3\over
  2\pi}\exp[-s(\sqrt{\sigma^2+p_3^2+gHn}+\lambda\mu')^2]. 
\end{equation}
Since we
assume that $\mu'^2$ and $gH$ are of the same order of magnitude and
large, $\mu'^2=O(\Lambda^2)\,\,gH=O(\Lambda^2)$, two terms, $n=0$ and
$n=1$, will make the main
contribution to the sum over $n$. The $n=0$ term is determined by the
integral, approximately equal to
\[
\int {dp_3\over
  2\pi}\exp[-s(\sqrt{\sigma^2+p_3^2}-\mu')^2]\approx {\mu'\over 2
  \sqrt{
  \pi s}\sqrt{\mu'^2-\sigma^2}}. 
\]
The $n=1$ term gives
\[
2\sum_{\lambda}\int {dp_3\over
  2\pi}\exp[-s(\sqrt{\sigma^2+p_3^2+gH}+\lambda\mu')^2]\approx
  {\mu'\over  \sqrt{
  \pi s}\sqrt{\mu'^2-\sigma^2-gH}},
\]
where we had introduced a factor two due to the two-fold degeneracy of
the $n=1$ Landau level in the spin variable. The above two terms with
$n=0$ and
$n=1$ should be summed up to give 
\[
V_{\rm eff 2}(\sigma)={gH\mu'\over 8\pi^2}\left({1\over \sqrt{\mu'^2
    -\sigma^2}}\ints_{1\over \mu'^2-\sigma^2}^{\infty} {ds\over
    s^2}+{2\over
  \sqrt{\mu'^2-gH-\sigma^2}}\ints_{1\over
  \mu'^2-\sigma^2-gH}^{\infty} {ds\over s^2}\right)
\]
\begin{equation}
={gH \mu'\over 8\pi^2}\left(\sqrt{\mu'^2-\sigma^2}+2\sqrt{
\mu'^2-gH-\sigma^2}\right).
\label{n01}
\end{equation}
The effective potential after summing the contributions $V_{\rm eff 1}$
and $V_{\rm eff 2}$ is finally estimated as follows:
\[
V_{\rm eff}(\sigma)={\mu'\over \pi^2}(\mu'^2-\sigma^2)^{3\over
2}+{3\sigma^4\over
    8\pi^2}\left[\left({2\Lambda^2\over\sigma^2}\right)^2-4{\Lambda^2
    \over
    \sigma^2}-\log {\sigma^2\over
    2\Lambda^2}+C_1\right]
\]
\begin{equation}
+{gH \mu'\over 8\pi^2}\left(\sqrt{\mu'^2-\sigma^2}+2\sqrt{
\mu'^2-gH-\sigma^2}\right)+{\sigma^2\over 4 G}+C.
\label{efftotal}
\end{equation} 
Next, we have to find the minimum point $\sigma_0$ of this effective
potential, i.e. to solve the stationarity equation $\partial V_{\rm
eff}/\partial \sigma^2=0$, which gives for $\sigma_0$ the equation
\[
-{3\mu'^2\over
2  \pi^2}-{3\Lambda^2\over 2\pi^2}-{3\sigma_0^2\over 4\pi^2}\log
  {\sigma_0^2\over2\Lambda^2}-{gH\over 16\pi^2}
\left(1+{2\mu'\over \sqrt{\mu'^2-gH}}\right)+{1\over
4G}+O(\sigma^2)=0. 
\]
It is seen that, contrary to the case with the pion condensate, there
are no 
large terms like $gH\log (gH/\sigma^2)$ in the above equation, and
hence, for large terms with $\Lambda_s$, it has a nontrivial solution
$\sigma_0\ne 0$ only if $\tilde g >1$, which is  the same condition
as for the zero field case (see, e.g., \cite{zhuang}),
\begin{equation}
\sigma_0=\sqrt{\tilde C
\Lambda}\exp\left(-\frac{\Lambda^2}{\sigma_0^2}(1-\frac
{1} {\tilde g})\right),
\label{pivii}
\end{equation}
where $\tilde C$ is a certain numerical constant of order unity.

Comparing the results of the particular
considerations {\bf a)} and {\bf b)},
the following preliminary conclusions can be made.
It is clear that at very small values of the
coupling constant $G$, i.e. at $\tilde g < 1$, and for $\mu'>0$,
both the isotopic and the chiral symmetry of the model 
(\ref{1}) are not broken down at $H=0$, and quarks are massless
particles (recall, we consider only the case  with zero current
quark mass). However,
if an external chromomagnetic field $H$ is 
present 
(and it is rather
strong in our analytical consideration), then the effective
potential (\ref{muuu}) of the system acquires a nontrivial global
minimum point which lies on the $\pi_1$-axis and has the form
$(\sigma=0,\pi_1=\pi_0\ne 0)$, where $\pi_0$ is the solution of the
stationarity equation (\ref{eqv}). This means that in this case a
nonzero pion condensate $\vev{\bar q \gamma_5\tau_1 q}=-\pi_0/2G$ is
generated by an external chromomagnetic field (chromomagnetic
catalysis
of the pion condensation), and the isotopic symmetry $ U_{I_3}(1)$
of the model is spontaneously broken
down (see the text after (\ref{1})). Moreover, since the expression
for the effective potential in 
this case resembles the effective potential in the 2-dimensional
Gross -- Neveu model, one can say that this effect, similar to the
magnetic catalysis phenomenon, is provided by the dimensional
reduction mechanism.   

Let us present
some numerical calculations in the next section, in order to support this conclusion.

\section{Numerical results}

\begin{figure}[htbp]
\includegraphics[width=200pt]{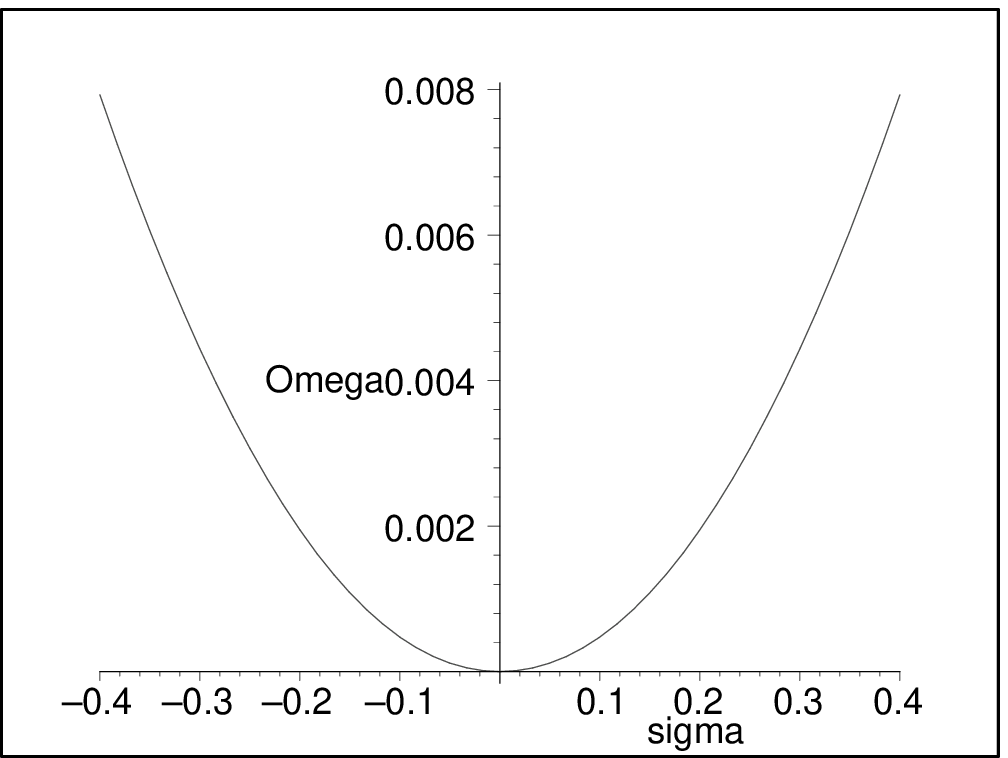}
\includegraphics[width=200pt]{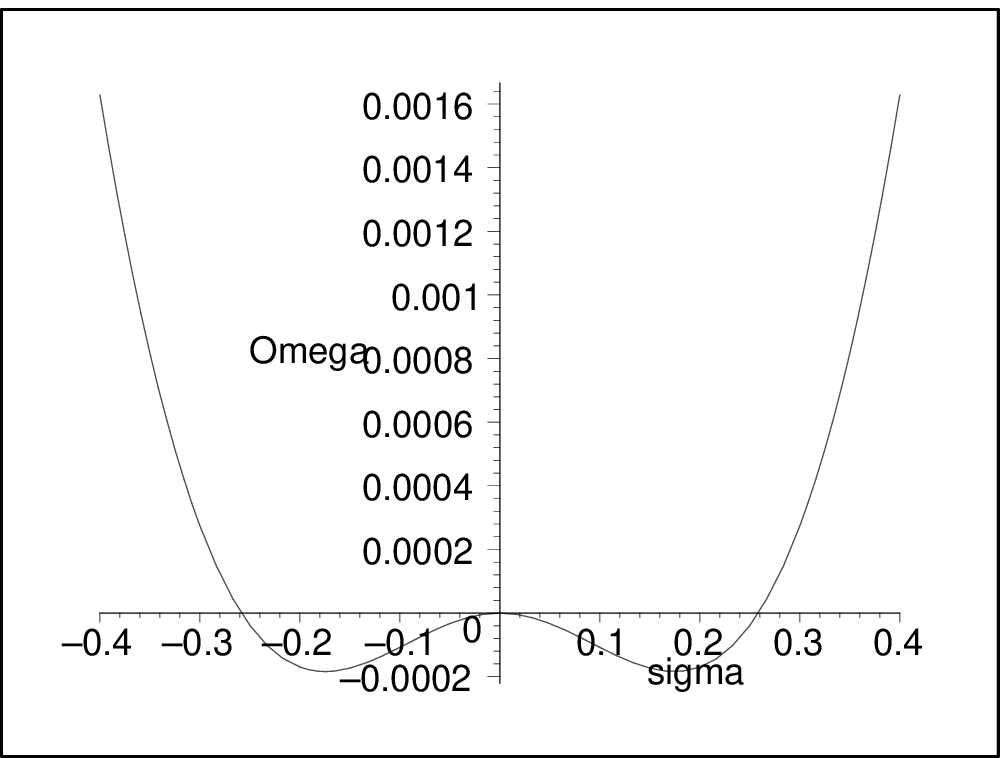}
\caption{Thermodynamic potential $\Omega=V_{\rm eff}(\sigma,0)$ at
$T=0$, $\mu'=0$, $\mu=0$ and $G=5.01$ GeV$^{-2}$ as a function of
  $\sigma$ (in GeV-units) for $\Lambda=0.165$ GeV -- left picture,
  and $\Lambda=0.65$ GeV -- right  picture.}
\label{fig:U2}
\end{figure}

\begin{figure}[htbp]
\includegraphics[width=300pt]{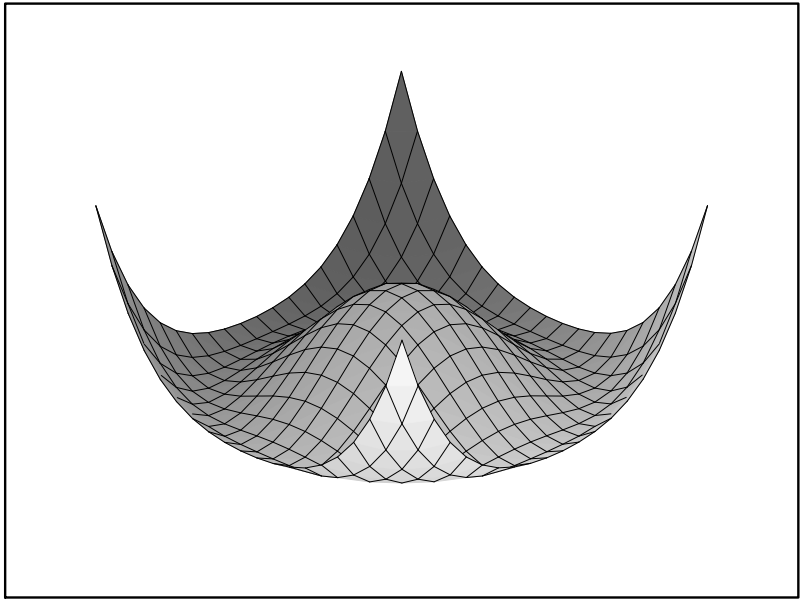}
\caption{Thermodynamic potential $\Omega=V_{\rm eff}(\sigma,\pi_1)$
at
  $T=0,\,\, gH=0,\,\, \mu'=0$ as a function of
  $\sigma,\,\,\pi_1$ for $\Lambda_p=0.65, \mu=0$. }
\label{fig:U3c}
\end{figure}

 Consider first the case of a zero external field $F_{\mu\nu}=0$, and
flavor symmetric medium $\mu'=0$. Moreover, the temperature and bare
quark mass $m$ are taken
to be zero throughout the present section. The numerical analysis of the
behavior of the effective potential (\ref{muuu}),
which is just the thermodynamic potential $\Omega$ 
for $T=0$,
was performed for the value of
the coupling constant $G=5.01$ GeV$^{-2}$, taken from
\cite{dekgk,zhuang}, and for the quark chemical
potential $\mu=0$ 
(see  Fig. \ref{fig:U2}).
In this case the quark energy
spectrum is 
symmetric with respect to $\sigma$ and $\pi_1$, i.e. the function
$V_{\rm eff}(\sigma,\pi_1)$  depends only on the single
variable $\sqrt{\sigma^2+\pi_1^2}$. Hence we draw the
effective potential as a function of only one variable, e.g.,
$\sigma$. In this figure, the  picture on the left corresponds to
the value of the cutoff parameter  $\Lambda =0.165$ GeV (in this case
$\tilde g=0.08$), whereas the
right picture corresponds to $\Lambda =0.65$ GeV ($\tilde g=1.29$). The
figure demonstrates the  general feature of the effective
potential at $\mu'=0$ and $gH=0$. Namely, if the effective coupling
constant $\tilde g$ is sufficiently small, $\tilde g<1$, then the
global minimum
point of the effective potential is at the origin, and the chiral
symmetry is not broken 
(left picture).
However, if $\tilde g>1$, the effective potential has a nontrivial
global minimum, and the chiral symmetry is spontaneously broken
down 
(right picture).
The dependence of $V_{\rm eff}(\sigma,\pi_1)$ on the two
variables 
$\sigma$ and $\pi_1$  ($\tilde g=1.29$) is depicted in Fig.
\ref{fig:U3c} for $\mu=0$.  The picture is evidently
symmetric, as it should be in this flavor symmetric case.

It is also necessary to note that if $\tilde g>1$, then at $gH=0$ and
$\mu'>0$ a nonzero pion condensate appears in the model (\ref{1})
\cite{dekgk,zhuang}. (In this case, the isotopic symmetry $U_{I_3}(1)$
is broken down).
However, if $\tilde g<1$, then at $gH=0$ and arbitrary values of
$\mu'$, including zero,  the symmetry of
the NJL model (\ref{1}) remains intact.

\begin{figure}[htbp]
\includegraphics[width=200pt]{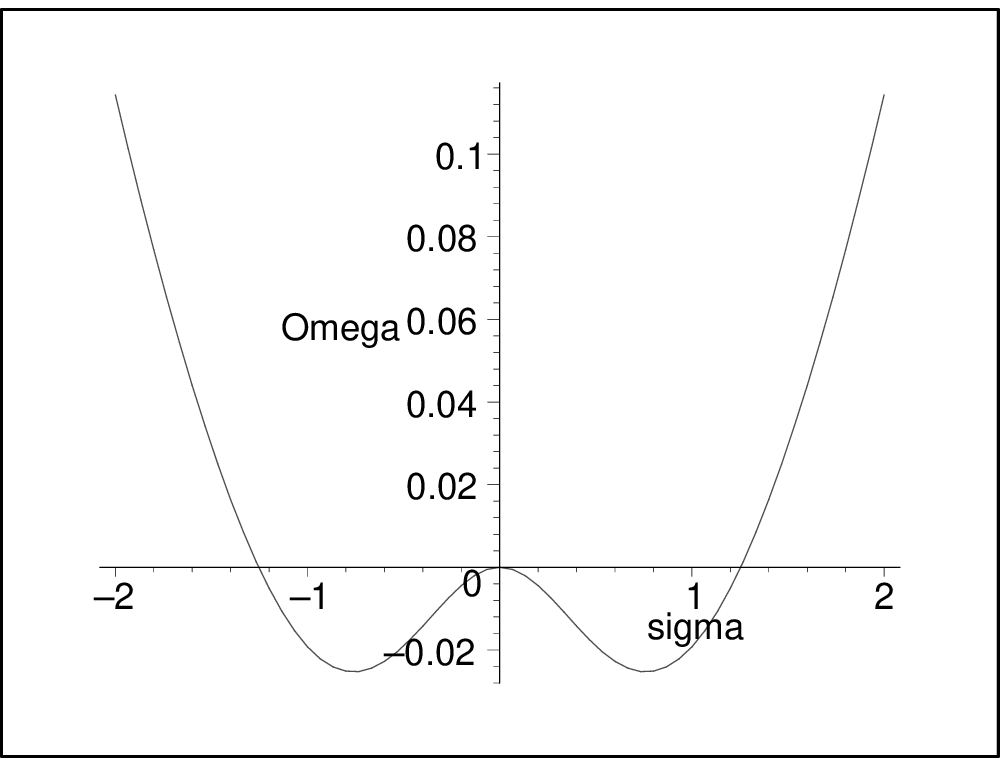}
\includegraphics[width=200pt]{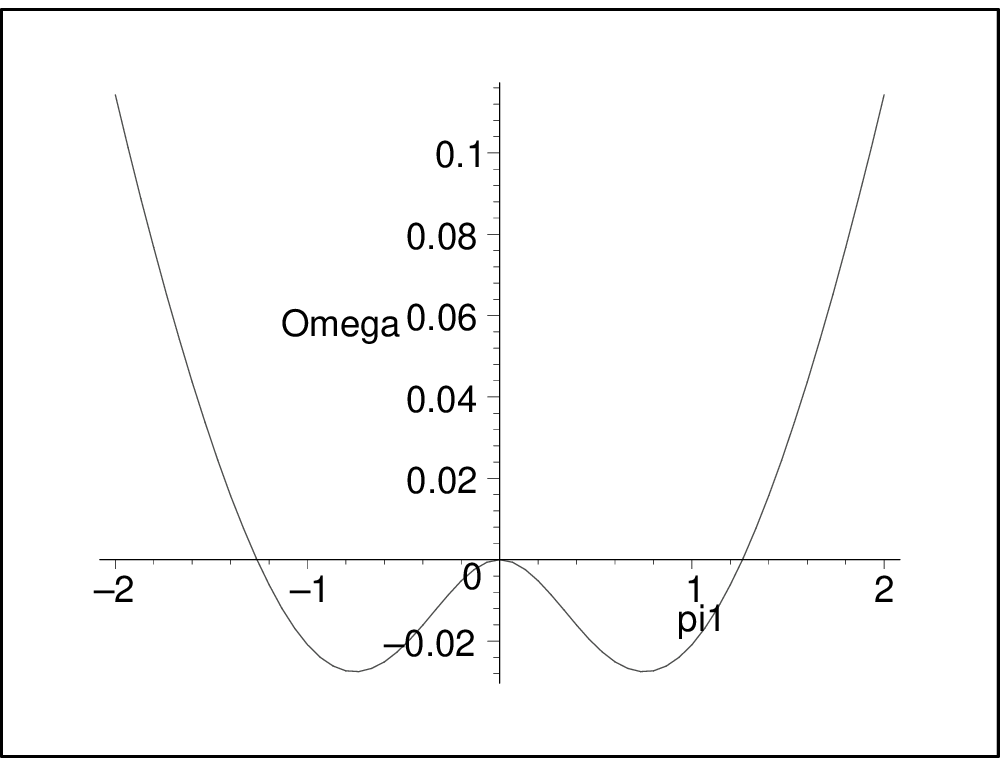}
\caption{Thermodynamic potential $\Omega=V_{\rm eff}(\sigma,\pi_1)$
at
  $T=0,\mu=0$, $gH=0.5$ GeV$^2$,   $\Lambda=0.65$ GeV,  $\tilde
  g=1.29$, $\mu'=0.15$ GeV as a function
  of $\sigma$ (in GeV-units) at $\pi_1=0$ -- left picture, and
  of $\pi_1$ (in GeV) at $\sigma=0$ -- right picture. }
\label{fig:U4a}
\end{figure}

\begin{figure}[htbp]
\includegraphics[width=200pt]{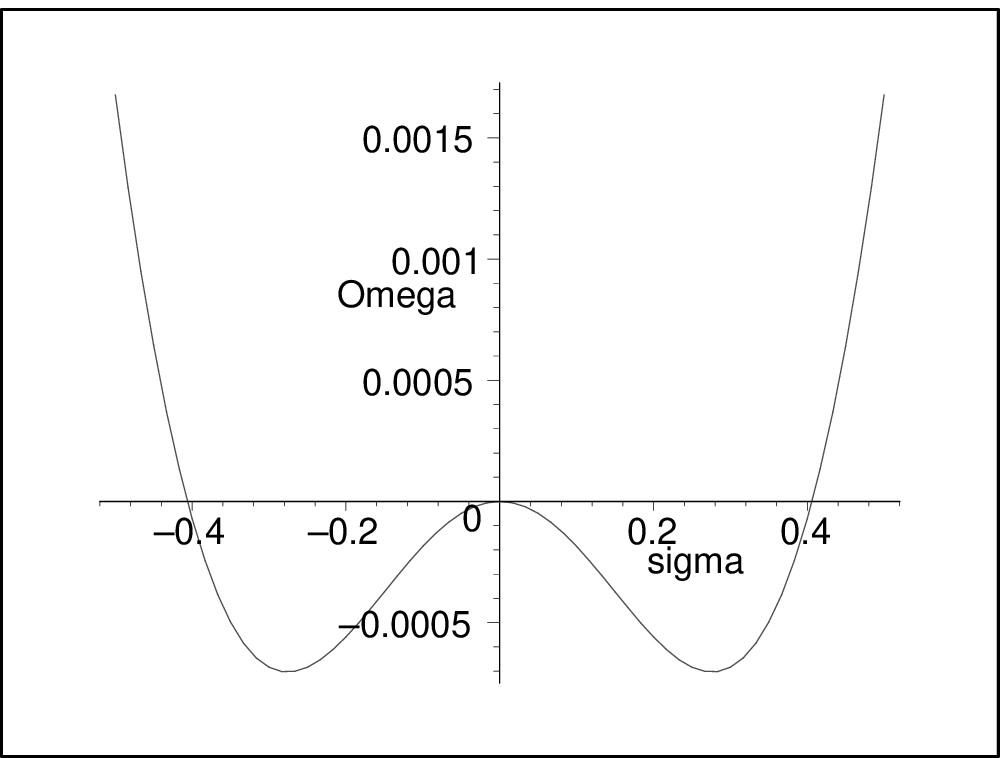}
\includegraphics[width=200pt]{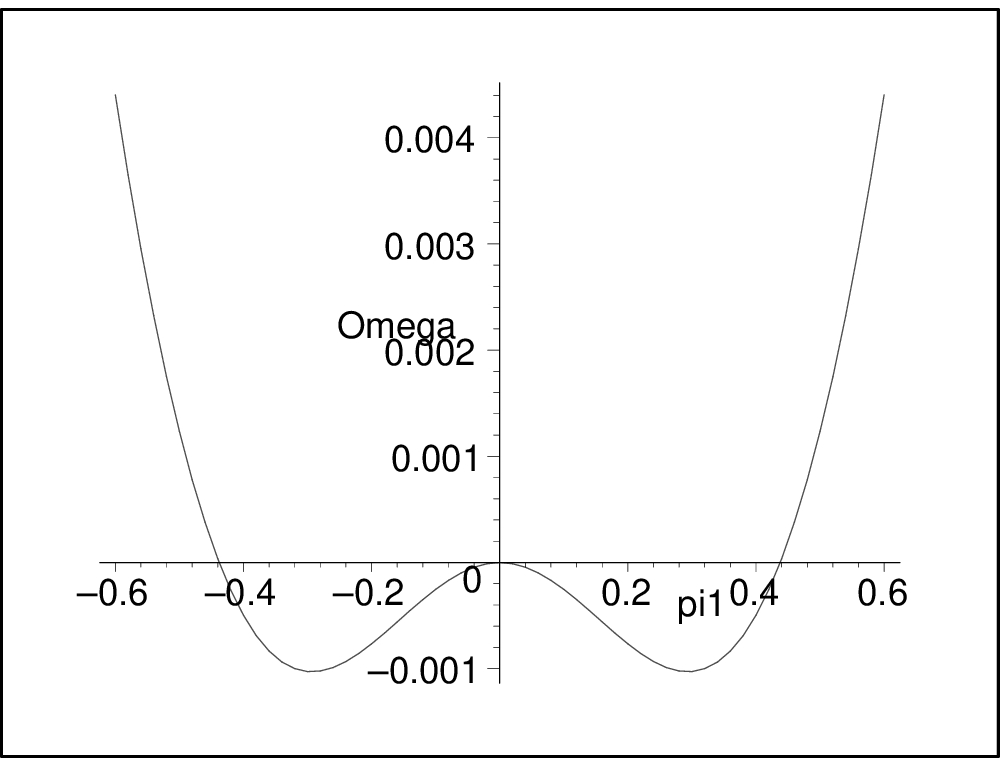}
\caption{Thermodynamic potential $\Omega=V_{\rm eff}(\sigma,\pi_1)$
at
  $T=0,\mu=0$, $gH=0.5$ GeV$^2$,   $\Lambda=0.46$ GeV,  $\tilde
  g=0.64$, $\mu'=0.1$ GeV as a function
  of $\sigma$ (in GeV-units) at $\pi_1=0$ -- left picture, and
  of $\pi_1$ (in GeV) at $\sigma=0$ -- right picture. }
\label{fig:U5b}
\end{figure}

\begin{figure}[htbp]
\includegraphics[width=200pt]{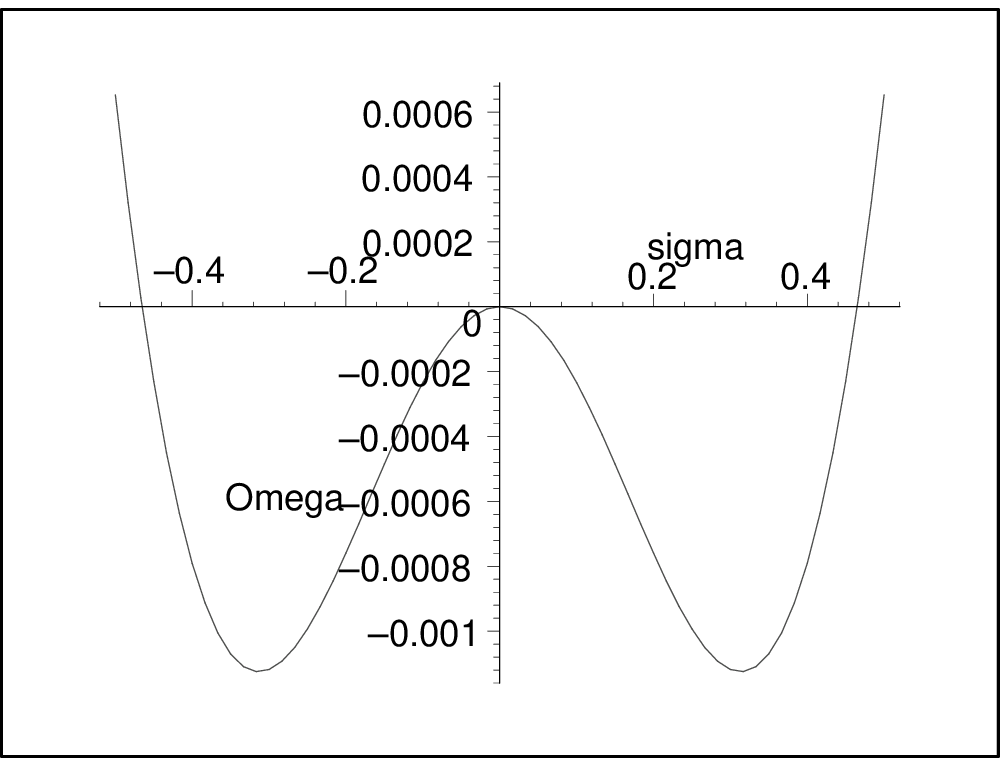}
\includegraphics[width=200pt]{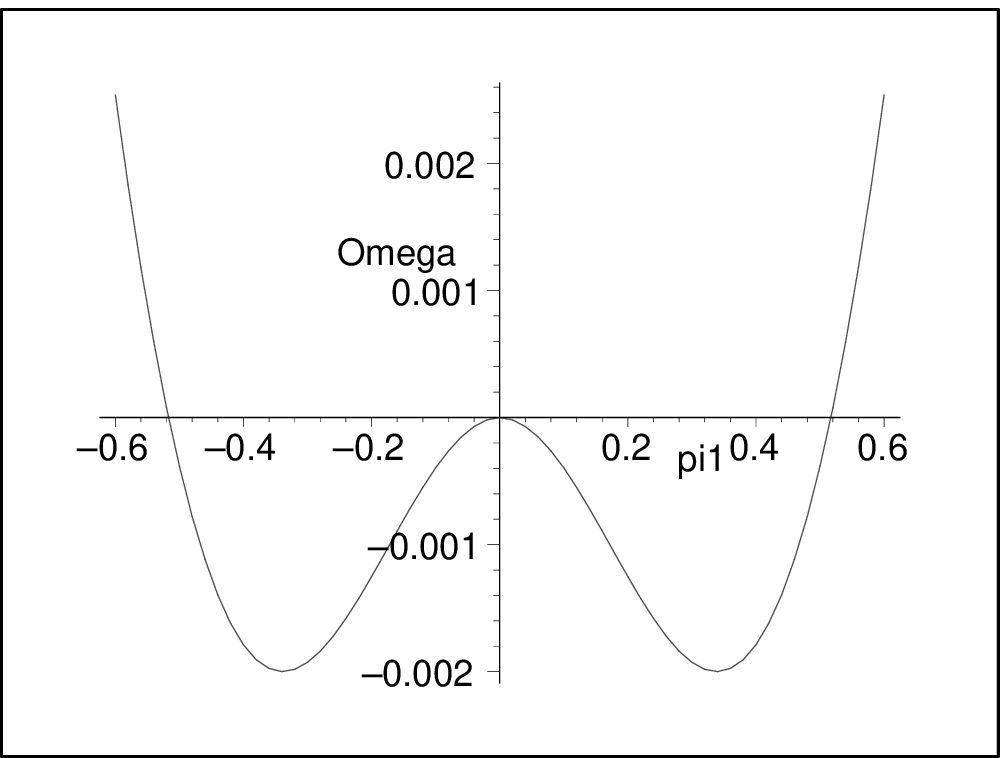}
\caption{Thermodynamic potential $\Omega=V_{\rm eff}(\sigma,\pi_1)$
at $T=0,\mu=0$, $gH=0.5$ GeV$^2$,   $\Lambda=0.46$ GeV,  $\tilde
g=0.64$, $\mu'=0.15$ GeV as a function
  of $\sigma$ (in GeV-units) at $\pi_1=0$ -- left picture, and
  of $\pi_1$ (in GeV) at $\sigma=0$ -- right picture.}
\label{fig:U5a}
\end{figure}

Now, in order to present numerical arguments in favour of the
chromomagnetic catalysis of pion
condensation, we will
study the effective potential $V_{\rm eff}(\sigma,\pi_1)$
(\ref{muuu})
at $\mu'>0$ 
and suppose that
$gH~=~0.5$~ GeV$^2$ (this value of $gH$ mimicks the nonzero QCD gluon
condensate \cite{nar}). Since the current quark mass $m$ is equal to
zero, the nontrivial stationary points of the function (\ref{muuu})
lie either on the $\sigma$ or on $\pi_1$ axis (see the remark following
formula (\ref{qqq})). Thus, in order to find the global
minimum point of the effective potential (\ref{muuu}), it is
enough to consider the behavior of $V_{\rm eff}(\sigma,\pi_1)$ along
the coordinate $\sigma$ and $\pi_1$-axis only. The corresponding
curves are presented in Fig.~\ref{fig:U4a} for $\mu'~=~0.15$~ GeV,
and physical values of $G=5.01$ GeV$^{-2}$ and $\Lambda =0.65$ GeV (in this case $\tilde
g=1.29$). 
They demonstrate that, although a nontrivial stationary
point 
of the thermodynamic potential $\Omega=V_{\rm eff}$ at $T=0$ 
does exist on the $\sigma$-axis  in a chromomagnetic 
field $gH$ (and this is just a saddle point in this
case),  
the value of the effective potential at the other stationary point,
 $\sigma =0,\pi_1=\pi_0\ne 0$, is slightly deeper.  Our calculations
 indicate that with growing $\mu'$
the difference between the minima of the curves for $V_{\rm
eff}(\pi_1)$
and $V_{\rm eff}(\sigma)$ increases, and the very minima get
deeper. The comparison of the results obtained with the known results at
zero background field, demonstrated that the background field deepens
the minimum for the pion condensate. Moreover, we formally considered
the interesting case of a smaller value of $\Lambda$, corresponding to
$\tilde g<1$. The calculated effective potential then has a minimum at a
nontrivial value of $\pi_0$ for weak coupling of quarks for $\tilde
g<1$, as well.     
The corresponding
curves are presented in Figs. \ref{fig:U5b}, \ref{fig:U5a} for  
different values of $\mu'=0.1$ GeV, $\mu'=0.15$ GeV, respectively,
and $G=5.01$ GeV$^{-2}$ and $\Lambda =0.46$ GeV (in this case $\tilde
g=0.64$).
\footnote{In contrast with the numerical results, in our analytical
investigations (see the previous section), the stationary point of
the
effective potential was not observed on the $\sigma$-axis at $\tilde
g<1$. This fact can be explained due to a different domain
for the variable $\mu'$ used in these cases. Indeed, it was supposed
in section II.B that $\mu'=O(\Lambda)$, whereas for the numerical
consideration, we used $\mu'\ll\Lambda$.}
Thus, only at this
point,
$\sigma =0$, $\pi_1=\pi_0\ne 0$, the global minimum of the
thermodynamic potential (\ref{muuu})
is observed, and the nonzero pion condensate $\vev{\bar q
\gamma_5\tau_1 q}=-\pi_0/2G$ is generated by the external
chromomagnetic field in the model.

\section{Conclusions}

We considered, in the framework of an NJL model, the effects of quark
and pion condensation in dense quark matter
with and without flavor asymmetry under the influence of an external
 chromomagnetic field modelling the gluon condensate
$\vev{F^a_{\mu\nu}F^{a\mu\nu}}$. 
The general conclusion is that the presence of the chromomagnetic
field catalyses 
the effect of quark or pion condensation, depending on the values of
the chemical potential $\mu'$. Indeed, 
when there is
no chromomagnetic field, no condensation of quarks or pions takes
place at low values of the effective
coupling 
(\ref{tg})
of quarks $\tilde g<1$. However,  when the chromomagnetic
background field is present, the pion (at $\mu'>0$) or quark
condesation (see the case $\mu'=0$ of Section II) take
place even for small values of $\tilde g<1$ (recall that in our
consideration the current quark mass $m$ was taken to be zero). 
If $\tilde g>1$, the role of the external chromomagnetic field $H$ is
to enhance the effect of 
pion condensation that appeares in the model at $\mu'>0$ already at
$H=0$.
For further investigations, it would be interesting to study 
the influence of an external
chromomagnetic field (the gluon condensate) on the pion condensation
phenomenon at nonzero current quark mass $m$ 
as well as to consider the chromomagnetic catalysis effect with
another
combinations of a background field, different from (\ref{13}).

\section{Acknowledgements}

One of the authors 
(V.Ch.Zh.) gratefully acknowledges the hospitality of
Prof. M.~Mueller-Preussker and his colleagues at the particle theory
group of the Humboldt University extended to him during his stay
there.
This work was partially supported by the Deutsche
Forschungsgemeinschaft under contract DFG 436 RUS 113/477/4 and RFBR
grant No. 05-02-16699.



\begin{thebibliography}{99}

\bibitem{7}
D. Ebert and M.K. Volkov, Yad. Fiz. {\bf 36}, 1265 (1982);
Z. Phys. {\bf C 16}, 205 (1983);
D. Ebert and H. Reinhardt, Nucl. Phys. {\bf B 271}, 188 (1986);
D. Ebert, H. Reinhardt and M.K. Volkov, Progr. Part. Nucl. Phys. {\bf
33}, 1 (1994).

\bibitem{kunihiro}
T. Hatsuda and T. Kunihiro, Phys. Rep. {\bf 247}, 221 (1994).

\bibitem{odin}
T. Inagaki, T. Muta and S.D. Odintsov,
Progr. Theor. Phys. Suppl. {\bf 127}, 93 (1997).

\bibitem{buballa}
M. Buballa, Phys. Rep. {\bf 407}, 205 (2005); G. Nardulli, Riv. Nuovo
Cim. {\bf 25N3}, 1 (2002).

\bibitem{klev}
V. Bernard, U.-G. Meissner and I. Zahed, Phys. Rev. {\bf D 36}, 819
(1987); M.~Asakawa and K.~Yazaki. Nucl. Phys. {\bf A 504},668 (1989);
J. H\"ufner, S.P. Klevansky and P.
Zhuang, Ann. Phys. {\bf 234}, 225 (1994).

\bibitem{klim}
A.S. Vshivtsev, V.Ch. Zhukovsky and K.G. Klimenko, 
JETP Lett. {\bf 64}, 338 (1996);
JETP {\bf 84}, 1047 (1997);
A.S. Vshivtsev, M.A. Vdovichenko and K.G. Klimenko,
JETP {\bf 87}, 229 (1998).

\bibitem{kl}
S.P. Klevansky and R.H. Lemmer, Phys. Rev. {\bf D 39}, 3478 (1989);
H. Suganuma and T. Tatsumi, Ann.of Phys.{\bf 208}, 470 (1991);
T. Inagaki, S.D. Odintsov and Yu.I. Shil'nov,
Int. J. Mod. Phys. {\bf A 14}, 481 (1999);
E. Gorbar, Phys. Lett. {\bf B 491}, 305 (2000).

\bibitem{ekvv}
M.A. Vdovichenko, A.S. Vshivtsev and K.G. Klimenko,
Phys. Atom. Nucl. {\bf 63}, 470 (2000);
D. Ebert et al, Phys. Rev. {\bf D 61}, 025005 (2000);
Phys. Atom. Nucl. {\bf 64}, 336 (2001);
Nucl. Phys. {\bf A 728}, 203 (2003).

\bibitem{2}
K.G. Klimenko, Teor. Mat. Fiz. {\bf 89}, 211 (1991);
{\bf 90}, 3 (1992);
Z. Phys. {\bf C 54}, 323 (1992);
V.P. Gusynin, V.A. Miransky and I.A. Shovkovy, Phys. Rev. Lett.
{\bf 73}, 3499 (1994); A.S. Vshivtsev, K.G. Klimenko and B.V.
Magnitsky, JETP Lett. {\bf 62}, 283 (1995).

\bibitem{gus}
V.P. Gusynin, V.A. Miransky and I.A. Shovkovy, Phys.Rev. {\bf D 52},
4747 (1996); Nucl. Phys. {\bf B 563}, 361 (1999); Phys. Lett.
{\bf B 349}, 477 (1995);
G.W. Semenoff, I.A. Shovkovy and L.C.R. Wijewardhana, Phys. Rev. {\bf
D 60}, 105024 (1999).

\bibitem{incera}
E.~J.~Ferrer and V.~de la Incera, Int. J. Mod. Phys.  {\bf 14}, 3963
(1999);
V.C. Zhukovsky et al, JETP Lett. {\bf 73}, 121 (2001); 
E.J. Ferrer, V.P. Gusynin and V. de la Incera, Eur. Phys. J. {\bf B
33}, 397 (2003); 
E. Elizalde, E.J. Ferrer, and V. de la Incera,
Phys. Rev. {\bf D 68}, 096004 (2003);
C.N. Leung and S.-Y. Wang, hep-ph/0510066;
C.G. Beneventano and E.M. Santangelo, hep-th/0511166.

\bibitem{gus2}
V.A. Miransky, Progr. Theor. Phys. Suppl.
{\bf 123}, 49 (1996); A.S. Vshivtsev et al,
Phys. Part. Nucl. {\bf 29}, 523 (1998);
V.P. Gusynin, Ukrainian J. Phys. {\bf 45}, 603 (2000); V. de la
Incera, hep-ph/0009303.

\bibitem{8}
D. Ebert and M.K. Volkov, Phys. Lett. {\bf B 272}, 86 (1991);
D. Ebert, Yu.L. Kalinovsky and M.K. Volkov, Phys. Lett. {\bf B 301},
 231 (1993).

\bibitem{klim2}
K.G. Klimenko, B.V. Magnitsky and A.S. Vshivtsev, Nuovo Cim. {\bf A
107}, 439 (1994);
Phys. Atom. Nucl. {\bf 57}, 2171 (1994);
Theor. Math. Phys. {\bf 101}, 1436 (1994);
I.A. Shovkovy and V.M. Turkowski, Phys. Lett. {\bf B 367}, 213
(1996).

\bibitem{zheb}
D. Ebert and V.Ch. Zhukovsky, Mod. Phys. Lett. {\bf A 12}, 2567
(1997).

\bibitem{saito}
K. Saito, K. Tsushima and A.W. Thomas, Mod. Phys. Lett. {\bf A 13},
769 (1998).

\bibitem{agas}
N.O. Agasian, B.O. Kerbikov and V.I. Shevchenko,
Phys. Rept. {\bf 320}, 131 (1999);
D. Ebert et al,
Phys. Rev. {\bf D 65} (2002) 054024; 
A. Iwazaki et al, Phys. Rev. {\bf D 71}, 034014 (2005);
D. Ebert, V.Ch. Zhukovsky and O.V. Tarasov, Phys. Rev. {\bf D 72},
096007 (2005).

\bibitem{ebklim}
V.Ch. Zhukovsky et al, JETP Lett. {\bf 74} (2001) 523;
hep-ph/0108185;
D.~Ebert, K.G.~Klimenko and H.~Toki,
Phys. Rev. {\bf D 64} (2001) 014038;
D. Ebert et al, Prog. Theor. Phys. {\bf 106}, 835 (2001).

\bibitem{ferrer}
E.~J.~Ferrer, V.~de la Incera, and C. Manuel, Phys. Rev. Lett. {\bf
95}, 152002 (2005);
hep-ph/0603233.

\bibitem{iwazaki}
A. Iwazaki, Phys. Rev. {\bf D 72}, 114003 (2005).

\bibitem{bub}
D. Toublan and J.B. Kogut, Phys. Lett. {\bf B 564}, 212
(2003); M. Buballa and M. Oertel, Nucl. Phys. {\bf A 703}, 770
(2002);
S. Lawley, W. Bentz and A.W. Thomas, Phys. Lett. {\bf B 632}, 495
(2006).

\bibitem{bed}
P.F. Bedaque, Nucl. Phys. {\bf A 697}, 569 (2002); H.J. Warringa, D.
Boer and J.O. Andersen, Phys. Rev. {\bf D 72},
014015 (2005).

\bibitem{barducci}
A. Barducci, R. Casalbuoni, G. Pettini, and L. Ravagli,
Phys. Lett. {\bf B 564}, 217 (2003); Phys. Rev. {\bf D 69},
096004 (2004).

\bibitem{zhuang}
L. He and P. Zhuang, Phys. Lett. {\bf B 615}, 93 (2005);
L. He, M. Jin, and P. Zhuang, Phys. Rev.  {\bf D 71},
116001 (2005);
hep-ph/0503249;
hep-ph/0604224.

\bibitem{dekgk} D.~Ebert and K.G.~Klimenko, J. Phys. {\bf G 32}, 599
(2006);
Eur. Phys. J. C 02527-5 (2006).

\bibitem{EbPerv}
D.~Ebert and V. N.~Pervushin, Teor. Mat. Fiz. {\bf 36}, 313 (1978);
D.~Ebert, Yu.L.~Kalinovsky, L.~M\"unchow and M.K.~Volkov, Int. J.
Mod.  Phys. {\bf A 8}, 1295 (1993).

\bibitem{Rho}
K. Langfeld and M. Rho, Nucl. Phys. {\bf A 660}, 475 (1999).

\bibitem{nar}
S. Narison, Phys. Lett. {\bf B 387}, 162 (1996); M. D'Elia, A. Di
Giacomo and E. Meggiolaro, Phys. Lett. {\bf B 408}, 315 (1997).

\bibitem{bieten}
M.A. Vdovichenko and A.K. Klimenko, JETP Lett.
{\bf 68}, 460 (1998); 
W. Bietenholz, A. Gfeller, and U.-J. Wiese, JHEP {\bf 10}, 018
(2003).


\end{thebibliography}
\end{document}